\documentclass[aps,prx,reprint,superscriptaddress, longbibliography]{revtex4-1}
\usepackage{H1}
\usepackage[pdftex,colorlinks=true]{hyperref}
\hypersetup{citecolor = blue, linktocpage=true}
\usepackage[normalem]{ulem}
\usepackage{color}
\usepackage[usenames,dvipsnames,svgnames,table]{xcolor}
\usepackage{mathtools}
\usepackage{enumerate}

\usepackage{tikz}
\usetikzlibrary{arrows}

\newcommand{\Tr}{ \mbox{Tr}}
\renewcommand{\Re}{ \mbox{Re}}

\newcommand{\mc}[1]{ { \mathcal {{#1}}}}

\newcommand{\half}{\frac{1}{2}}
\renewcommand{\ket}[1]{\left|#1\right\rangle}

\begin{document}
\title{Asymmetric butterfly velocities in Hamiltonian and circuit models}
\author{Charles Stahl}
\affiliation{\mbox{Department of Physics, Princeton University, Princeton, NJ 08544, USA}}
\affiliation{\mbox{Department of Applied Mathematics and Theoretical Physics, University of Cambridge, Cambridge, UK}}
\author{Vedika Khemani}
\affiliation{\mbox{Department of Physics, Harvard University, Cambridge, MA 02138, USA}}
\author{David A. Huse}
\affiliation{\mbox{Department of Physics, Princeton University, Princeton, NJ 08544, USA}}

\begin{abstract}
The butterfly velocity $v_B$ has been proposed as a characteristic velocity for information propagation in local systems. It can be measured by the ballistic spreading of local operators in time (or, equivalently, by out-of-time-ordered commutators). In general, this velocity can depend on the direction of spreading and, indeed, the asymmetry between different directions can be made arbitrarily large using arbitrarily deep quantum circuits. Nevertheless, in all examples of local time-independent Hamiltonians that have been examined thus far, this velocity is independent of the direction of information propagation. In this work, we present two models with asymmetric $v_B$. The first is a time-independent Hamiltonian in one dimension with local, 3-site interactions. The second is a class of local unitary circuits, which we call $n$-staircases, where $n$ serves as a tunable parameter interpolating from $n=1$ with symmetric spreading to $n=\infty$ with completely chiral information propagation.  
\end{abstract}

\maketitle

\section{Introduction}
Understanding the dynamics of isolated quantum systems is a topic of fundamental interest. One central question is how isolated systems undergoing unitary time evolution are able to bring themselves to local thermal equilibrium under their own dynamics \cite{Deutsch,Srednicki, Rigol}. Indeed, while all quantum information is always preserved under unitary evolution, it can get ``scrambled" in highly non-local, experimentally inaccessible degrees of freedom - leading to an effective decoherence that can bring local subsystems to thermal equilibrium~\cite{Nandkishore14}. 

A useful window into the scrambling process comes from studying the spreading of initially local perturbations under time evolution. Under Heisenberg evolution, a local operator $O_0$ evolves into $O_0(t) = U^\dagger(t) O_0 U(t)$ with support on a spatial region that grows with time. We focus here on chaotic systems with ballistic information spreading at a ``butterfly speed" $v_B(\hat{\vec{n}})$ which may, in principle, depend on the direction of propagation $\hat{\vec{n}}$. The ``footprint" of the spreading operator defines an effective ``light-cone" bounded by operator fronts propagating in different directions with speed $v_B(\hat{\vec{n}})$. One well-studied diagnostic of this operator spreading is the out-of-time ordered commutator (OTOC)~\cite{Larkinotoc, chaosbound, ShenkerStanfordButterfly}, discussed in Section~\ref{sec:HamOTOC}. The primary goal of this paper is to explore models with asymmetric spreading of quantum information in different directions \emph{i.e.} models where $v_B$ demonstrably depends on the direction of propagation.

\begin{figure}
	\includegraphics[width=\columnwidth]{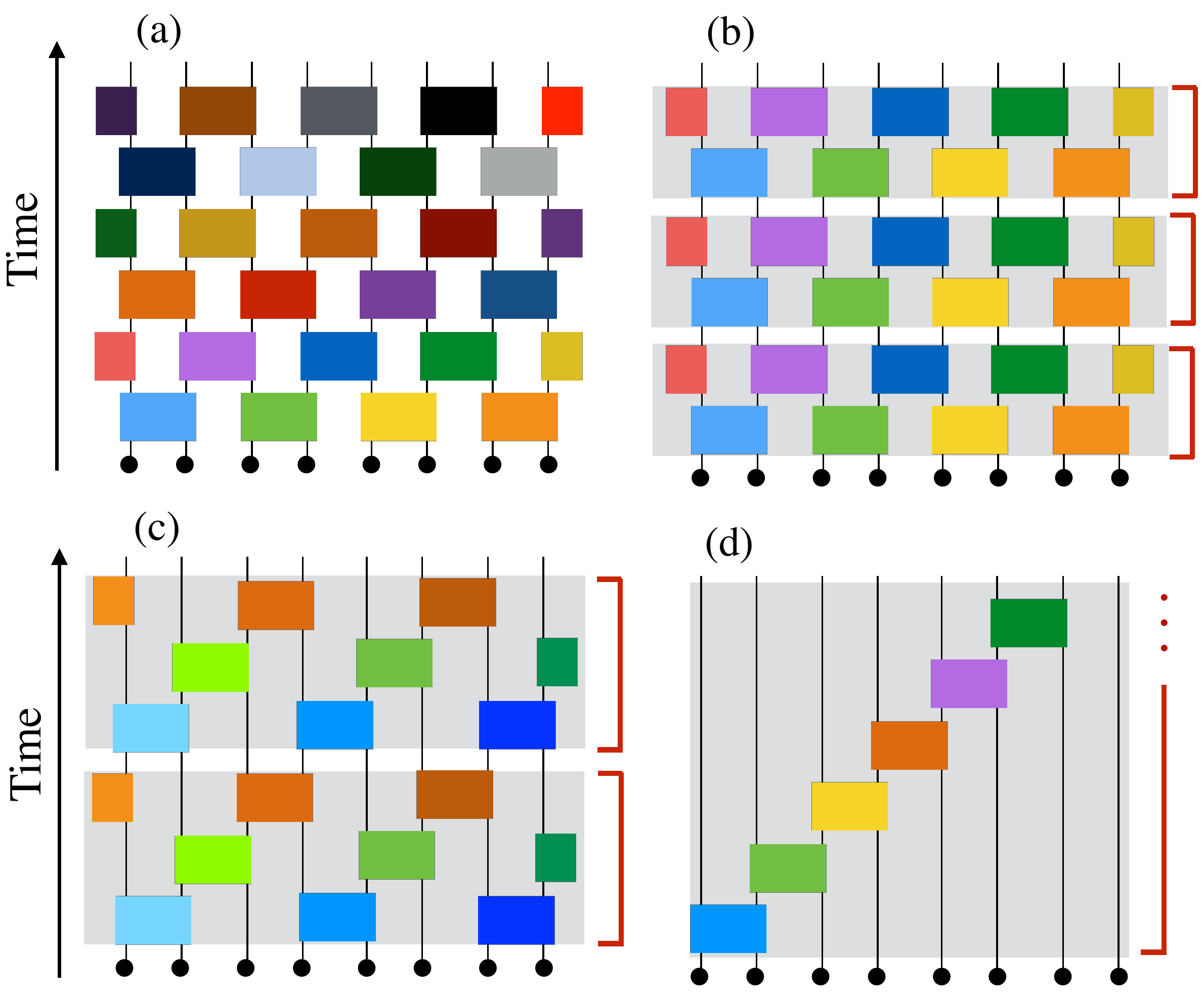}
	\caption{Random unitary circuits in one dimension. Each gate (rectangular box) is a unitary operator that acts on two adjacent sites and is drawn randomly from some ensemble, such as the uniform Haar measure. The different colors of gates represent different randomly drawn gates. The gates in circuit (a) are random in space and time, while the circuits in (b)-(d) are Floquet circuits which are periodic in time. The ``depth" of the Floquet circuits is set by the time-period which is, respectively, (b) two, (c) three, and (d) $L$ for a chain of length $L$. The Floquet circuit geometries in (c), (d) are chosen to give asymmetric speeds for information transfer to the left and right, with (d) being a completely unidirectional chiral circuit. 
	}
	\label{fig:circuits}
\end{figure}

Recently, a great deal of insight into the dynamics of operator spreading and quantum entanglement in chaotic systems has been gained by studying various minimally structured coarse-grained models whose time evolution is generated by random unitary circuits \cite{FawziScrambling,HosurYoshida,AdamCircuit1,opspreadAdam, opspreadCurt, TiborCons, KhemaniCons, ChanFloquetRC}.
Such models are analytically tractable, by design, and are constrained only by locality, unitarity and a few local conservation laws. 
A central assumption, borne out by these studies, is that the time evolution in chaotic systems looks essentially random, so that these ingredients are sufficient for capturing several universal features of the dynamics of thermalizing systems. Several variations have been studied, including unitary circuits that are either random~\cite{AdamCircuit1,opspreadAdam, opspreadCurt, TiborCons, KhemaniCons} or periodic~\cite{ChanFloquetRC} in time, the latter called ``Floquet" circuits. Circuits with unitary gates drawn randomly from various ensembles have been considered: for example, uniformly from the Haar measure, or randomly from the Clifford group, or random gates subject to various local conservation laws~\cite{KhemaniCons, TiborCons}. Figure~\ref{fig:circuits} depicts some of these cases. In addition, one can also consider circuits with random spatial architectures, so that random unitary gates are dropped at random spatial locations at every time step~\cite{AdamCircuit1, NahumRuhmanHuse}.    

In this vein, models of unitary circuits with spatially \emph{asymmetric} information propagation are straightforward to realize. These include the ``glider" Clifford circuits~\cite{Gutschow2009, Gutschow2010} and the translation operator in one dimension~\cite{PoChiralFloquet}, for which information propagation is completely unidirectional. The circuit architectures of these models is similar to Fig.~\ref{fig:circuits}(d) where, in the periodic Floquet setting, unidirectional or chiral information propagation requires circuits of infinite depth (time period) scaling with system size $~L$~\footnote{It is possible to get chiral information transport on the edge of a two-dimensional model using only a finite-depth Floquet circuit~\cite{RudnerFloquetEdge, PoChiralFloquet}.}. On the other hand, a finite but unequal ratio of left and right propagation speeds can be achieved using Floquet circuits of finite depth; this is clear from Fig~\ref{fig:circuits}(c) which depicts a period three Floquet circuit built from length three ``staircases" in which operators spread to the right twice as fast as they spread to the left, on average. 

These examples suggest that asymmetric information propagation can be realized quite generally, and \emph{a priori}, should be realizable even in local, time-independent Hamiltonian systems. Nevertheless, to the best of our knowledge, almost all examples of local time-independent Hamiltonians that have been examined thus far show symmetric propagation. One of our main results is to construct a time-independent Hamiltonian with local three-site interactions which displays asymmetric information propagation in the left and right directions (Section~\ref{sec:ham}). We quantify the asymmetry by measuring the left and right butterfly speeds, $v_{B, l}$ and  $v_{B,r}$, using various measures of operator spreading. 
Our construction is inspired by results from asymmetric circuit models, and can be easily generalized to realize greater degrees of asymmetry by making the Hamiltonian more non-local. In addition, in Section~\ref{sec:circ}, we also introduce and analyze a class of random circuit models,  called $n$-staircase circuits, in which the asymmetry in butterfly speeds can be tuned by tuning $n$. In such circuits, the ``entanglement generation function" governing the entanglement and operator dynamics, introduced in Ref.~\cite{Jonay}, can be tuned to have any shape (consistent with the convexity conditions discussed in \cite{Jonay}). 

\begin{figure}
	\includegraphics[width=\columnwidth]{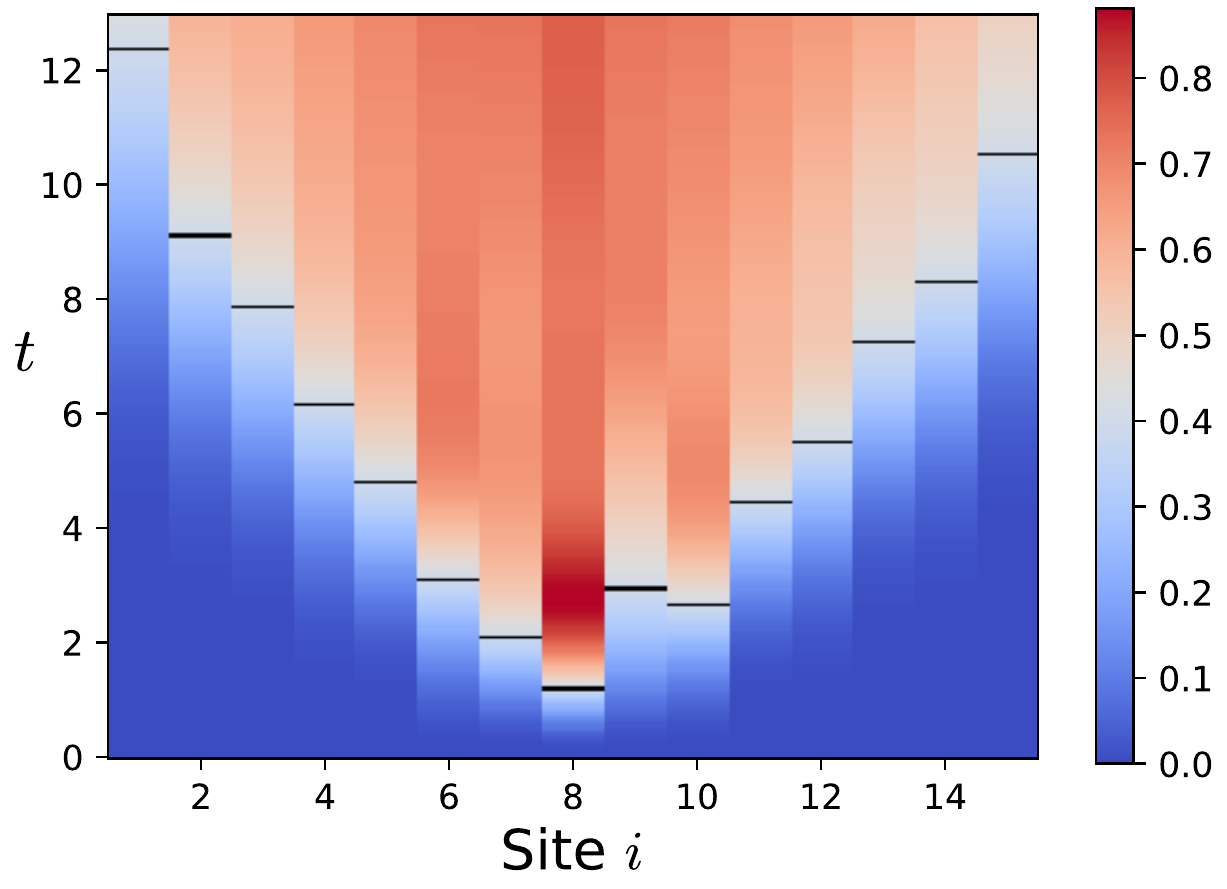}
	\caption{Evolution of the OTOC $C(i,t)$ for an operator initially located at the center of the chain for the model in Equation~\ref{eq:Hammodel}. Data is averaged over 100 disorder samples at $L=15, h=0.35$ with the initial operator at the central site. The bars indicate the time at which the OTOC passes 0.4, to emphasize the asymmetry.}
	\label{fig:colorplot}
\end{figure}

%\tableofcontents

\section{Local Time-Indepedent Hamiltonians with Asymmetric butterfly speeds}
\label{sec:ham}
In this section, we will construct an explicit example of a time-independent local Hamiltonian with asymmetric information propagation. We will work with spin $1/2$ models on a one dimensional lattice with $L$ sites and open boundary conditions, and consider spatially local Hamiltonians of finite range $n$:
\begin{align}
H = \sum_{i=1}^{L-n+1}H_n^{(i)},
\label{eqn:chain}
\end{align}
where $H_n^{(i)}$ is an $n$-site Hamiltonian acting on sites $i$ through $i+n-1$. The degree of asymmetry between the left and right butterfly speeds can by varied by varying $n$. 

Our strategy will be to construct the ``building blocks" $H_n^{(i)}$ to \emph{individually} show some asymmetry in information transfer. Note that $n=2$ will not suffice for this purpose, because 2-site Hamiltonians are always symmetric with respect to their operator dynamics. Unitarity preserves the total amount of information, so if the 2-site Hamiltonian moves some weight from the first site to the second, it must move an equal amount from the second to the first. On the other hand, 3-site Hamiltonians do not have this constraint, and can have asymmetric dynamics. Thus, our minimal example of a Hamiltonian showing asymmetric dynamics will have three-site interactions. 

Instead of looking directly for an asymmetric $H_3$, we can find a unitary operator $U_3$ with the desired dynamics. We can then invert that to obtain $H_3$ such that $U_3=e^{-iH_3}$. One such asymmetric unitary operator is the 3-site cyclic permutation operator $S_{123}$, whose action is defined by:
\begin{align}
S_{123}\ket{\alpha\beta\gamma} =\ket{\gamma\alpha\beta}, \label{eqn:condition}
\end{align}
where $\ket{\alpha\beta\gamma}$ is a product state with state $\ket{\alpha}$ on site 1, etc. This operator can transport a state from site 3 to site 1 in one step, but it takes two applications to move a state from site 1 to site 3. One way to build the three site swap gate is out of 2-site SWAP gates $S_{123} = S_{23}S_{12}$. Each 2-site SWAP interchanges two states, so the action is 
\begin{align}
S_{12}S_{23}\ket{\alpha\beta\gamma} &= S_{12}\ket{\alpha\gamma\beta} = 
\ket{\gamma\alpha\beta}\nonumber\\
&= S_{123}\ket{\alpha\beta\gamma}.
\end{align}
This construction can be easily extended to $n$ sites to create an $n$-site cyclic permutation gate $S_{12\cdots n}$ which can be written as a series of overlapping 2-site SWAP gates. In Section~\ref{sec:circ} we will use this architecture, in a very different system, to build our asymmetric circuits. 

Although we will eventually turn $S_{123}$ into a time-independent Hamiltonian, one could instead consider a period three Floquet unitary built from $S_{123}$ gates acting regularly on a spin chain. 
The Floquet unitary would be
$U = \left(\prod_i S_{123}^{(3i+1)}\right) \left(\prod_i S_{123}^{(3i+2)}\right) \left(\prod_i S_{123}^{(3i)}\right) $
so that the three-site gates at a given time step are non-overlapping in space, but they overlap between time-steps in a three-site generalization of the ``brickwork circuit" shown in Figs~\ref{fig:circuits}(a,b). It is straightforward to show that rewriting the $S_{123}$ gates as two-site SWAP gates gives a Floquet circuit whose architecture is equivalent to the asymmetric Floquet circuit in Fig~\ref{fig:circuits}(c). This clearly results in asymmetric spreading, so we are on the right track.
Again, for $n$-site permutations gates, these will be generalized to Floquet circuits of period $n$ and increasing degrees of asymmetry. 

We can now use $S_{123}$ to obtain the desired $H_3$. There are many ways to construct this Hamiltonian, from directly taking the matrix logarithm to analyzing eigenstates. But the simplest way to do this is to note that exchanging any two site indices gives $S_{123}^{-1}$, while overall SU(2) rotations leave the gate unchanged. This means $H_3$ should be antisymmetric with respect to exchanging site indices, and symmetric with respect to SU(2). Therefore $H_3$ is proportional the triple product of the spin on three sites, $H_3 = {\bf S}_1\cdot({\bf S}_{2}\times {\bf S}_{3})$. Putting it together, a candidate Hamiltonian on the full chain is then
\begin{align}
H = \sum_{i=1}^{L-2}{\bf S}_i\cdot({\bf S}_{i+1}\times {\bf S}_{i+2}),
\label{eq:Hamtriple}
\end{align}
where $\bf{S}_i$ are spin 1/2 operators on site $i$. Similarly, more non-local Hamiltonians with greater degrees of asymmetry can be constructed by defining $H_n$ as the matrix logarithm of $S_{12\cdots n}$

In the following subsections, we will numerically analyse the dynamics in model \eqref{eq:Hamtriple} in more detail, confirming our expectation that this model shows asymmetric operator spreading. In particular, because states and operators evolve in opposite directions, we expect the butterfly velocity in the right direction to be larger for \eqref{eq:Hamtriple} (See Figure~\ref{fig:colorplot}). We also note that the use of 3-site terms has some further consequences. For example, first order perturbation theory will connect site 1 to sites 2 and 3, while second order perturbations connect site 1 to sites 4 and 5. At early time sites 2 will behave the same as site 3, etc., leading to even-odd effects in the spreading. We will correct for this by only looking at alternate sites for each analysis. 

\subsection{Spectral Degeneracies, and Choice of Model}
\label{sec:HamChoice}
\begin{figure}
	\includegraphics[width=\columnwidth]{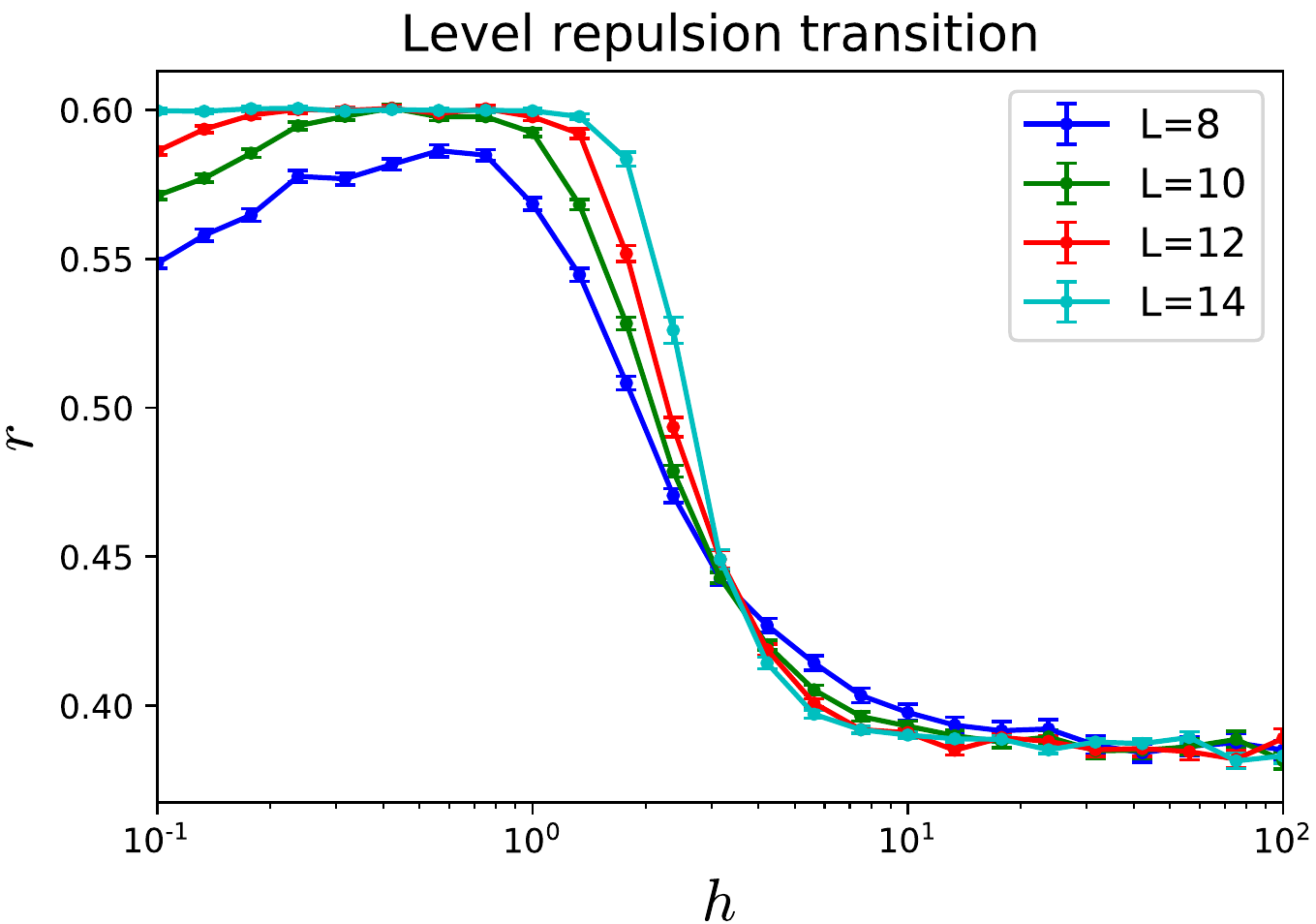}
	\caption{Phase transition for the model \eqref{eq:Hammodel}, with the level repulsion parameter $ r$  plotted against the strength of randomness $h$. For small $h$, the r-ratio $ r  \sim 0.6$, appropriate to a chaotic system with GUE statistics, while for large $h$, the r-ratio flows towards the Poisson value of $0.386$ with increasing $L$, characteristic of localization.}
	\label{fig:levelrepultrans}
\end{figure}

While we have confirmed that \eqref{eq:Hamtriple} shows asymmetric operator spreading, the dynamics in this model suffers from various non-generic peculiarities due to the presence of several additional symmetries, with one symptom being an exponentially large degeneracy of energy eigenvalues at $E=0$. Thus, we now introduce a perturbed version of \eqref{eq:Hamtriple} which retains the asymmetric information transfer but is more generic.

First, to see the presence of the exponentially large zero-energy degeneracy, note that \eqref{eq:Hamtriple} anticommutes with inversion symmetry $\mathcal{I}$ so that the eigenenergies come in $\pm E$ pairs for $E\neq0$. It is straightforward to show that in the presence of operators $R$ such that $\{H, R\}=0$, the degeneracy of the zero-energy manifold is lower-bounded by $\mbox{Tr}(R)$~\cite{IadecolaFSUSY}. The inversion operator $\mathcal{I}$ is one such operator with $\mbox{Tr}(\mathcal{I}) \sim 2^{L/2}$, partially explaining the zero-energy degeneracy. In fact, for even length chains, the degeneracy is even larger than $\mbox{Tr}(\mathcal{I})$ because of the presence of the additional $SU(2)$ symmetry. If we break the $SU(2)$ symmetry down to $U(1)$, say by adding a uniform field in the $Z$ direction, then $\mathcal{I}$ no longer anticommutes with $H$, but $R = P_x \mathcal I $ does, where $P_x = \prod_i S_i^x$. In this case, $\mbox{Tr}(R)$ gives the zero energy degeneracy for both even and odd $L$. Finally, we can get rid of all degeneracies by adding a random field in the $Z$ direction, so that the Hamiltonian is \begin{align}
H = \sum_{i=1}^{L-2}{\bf S}_i\cdot({\bf S}_{i+1}\times {\bf S}_{i+2}) + 
	\sum_{i=1}^{L}h_iS_i^z,
\label{eq:Hammodel}
\end{align}
where each $h_i$ has a uniform probability distribution on $[-h,h]$. This model, with a suitably chosen $h$ will be our model of choice. 

To choose an apporpriate $h$, note that while the model above \eqref{eq:Hammodel} certainly gets rid of various peculiarities present in the dynamics of \eqref{eq:Hamtriple}, the random field introduces the possibility of many-body localization for large enough $h$~\cite{Nandkishore14}. One widely used diagnostic of localization is the level spacing ratio, defined as $r_n = {\rm{min}}(\Delta E_{n+1}/\Delta E_n,\Delta E_{n}/\Delta E_{n+1})$ where $\Delta E_n = E_n - E_{n-1}$ and $E_n$ is the $n$th energy eigenvalue~\cite{OganesyanHuse, PalHuse}. This parameter is a probe of the level repulsion in the eigenspectrum, and the spectrally averaged $r_n$, $r$, flows towards the GUE value $0.6$, while it flows towards the Poisson value $0.386$ in a localized system~\cite{Atas}. 
In Fig.~\ref{fig:levelrepultrans}, we plot $r$ as a function of $h$ averaged over the middle third of the spectrum and 100-1000 disorder realizations depending on the system size. This number of disorder samples was chosen to reach a small enough error. We see a transition to a localized phase near $h \sim 3$. To steer clear of both the $h=0$ and large $h$ limits, we work with $h \sim 0.35$ for the balance of this paper. This field is also small enough that the dynamics are still dominated by the triple-product term, leading to the desired asymmetry (see Fig.~\ref{fig:colorplot}).

\subsection{Asymmetric butterfly speeds from left/right operator weights}
\label{sec:HamRight}
We now turn to an analysis of the asymmetric butterfly speeds in \eqref{eq:Hammodel} measured through the spreading of local operators. For spreading in the right (forward) direction, the initial operator is placed on site 1, while for spreading in the left (backwards direction), the initial operator is placed at site $L$. We will quantify the asymmetry in spreading speeds using two metrics: the right/left weights, defined below, and the OTOC. 

\begin{figure}
	\includegraphics[width=\columnwidth]{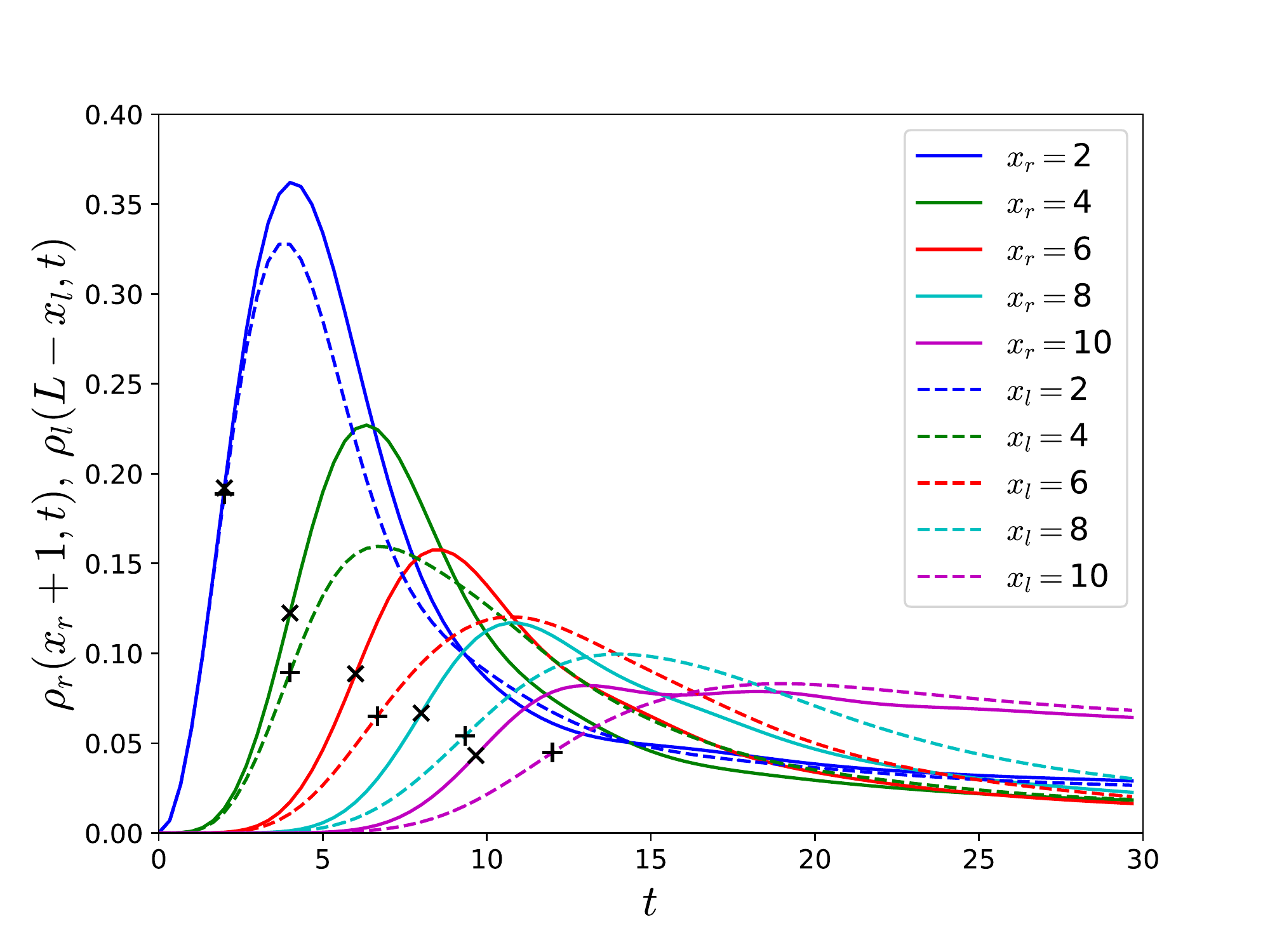}
	\includegraphics[width=\columnwidth]{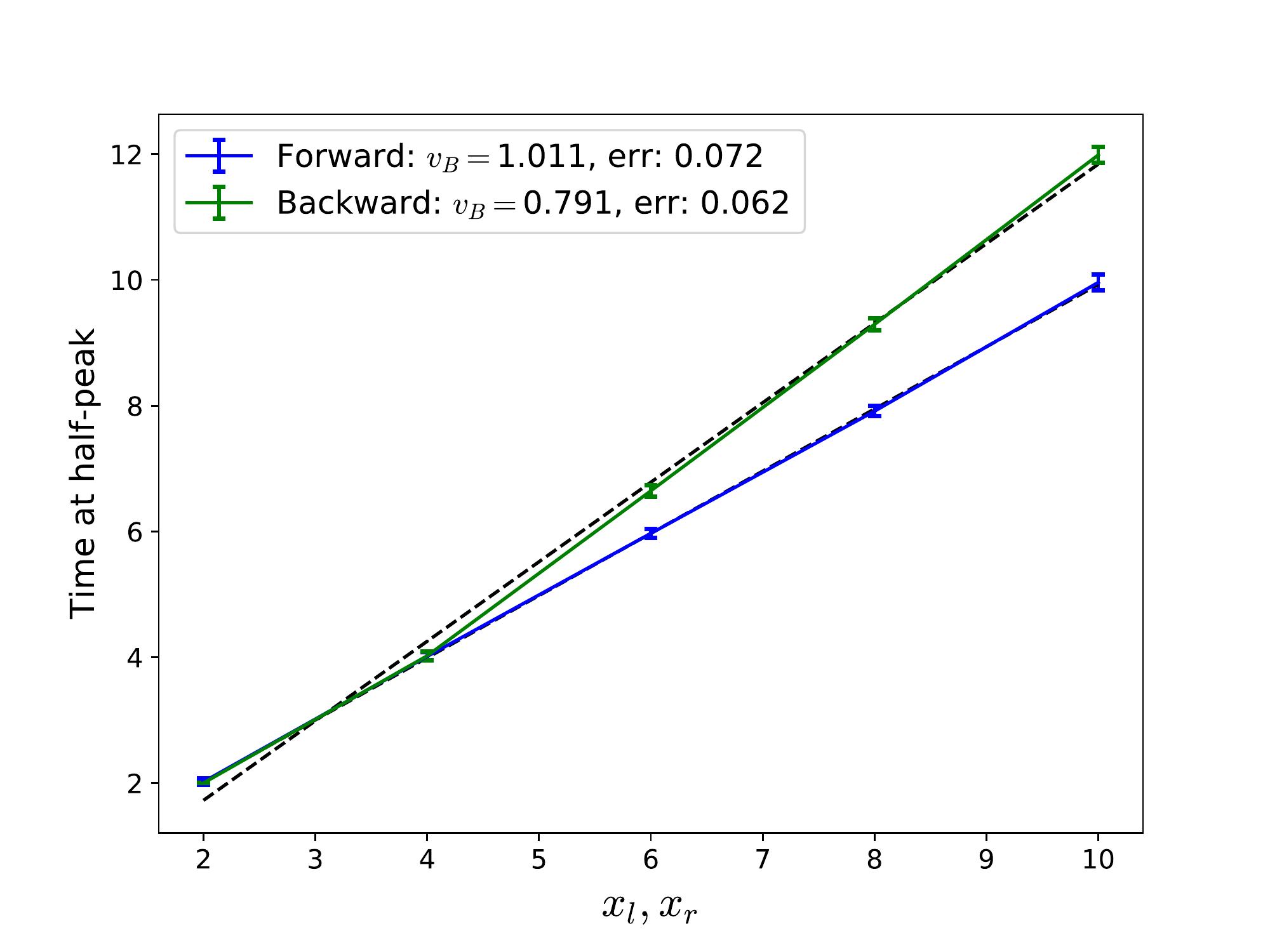}
	\caption{(top) Right weights (solid lines) and left weights (dashed lines) at even distances $x$ from the starting site for $L=13$, $h=0.35$, averaged over 100 disorder realizations. For $\rho_r(1+x,t)$, the initial operator is $\sigma^z_1$ while for $\rho_l(L-x,t)$ the initial operator is $\sigma^z_{L}$. The peaks travel ballistically, and curves for $x=0$ and $x=L-1=12$ are excluded for clarity. The symbols $\times/+$ mark the time at which the right/left weights reach half their maximum peak height for a given distance $x$. The right weight peaks earlier at later times, signifying a faster butterfly velocity in the forward direction.
	(bottom) Time of half-peak for right/left weights vs. distance $x$. The linear fit confirms ballistic propagation of the left and right fronts. Since this is plot of time as a function of distance, the larger slope in the left weight means that $v_B$ is larger for propagation to the right. 
	}
	\label{fig:Rweightpeakshape}
\end{figure}

To define the right and left weights, note that in a spin 1/2 system of length $L$, a complete orthonormal basis for all operators is provided by the $4^L$ ``Pauli strings" $\mathcal{S}$ which are products of distinct Pauli operators $\{1, \sigma^x, \sigma^y, \sigma^z \}$ on each site:
$O(t) = \sum_{\mathcal{S}} a_\mathcal{S}(t) \mathcal{S}$. Unitarity enforces that the norm of the operator is constant for all times, which means $\sum_{\mathcal{S}} |a_\mathcal{S}(t)|^2 =1$ for a normalized operator. 
An initially local operator, at early times, consists only of strings $\mathcal{S}$ that are the local identity everywhere except near the starting position. But, with time, the operator weight spreads to longer Pauli strings, containing
non-identity local operators at sites out to fronts whose distance from the origin grows ballistically with time. It is this operator growth that we will be measuring in both the right and left directions. On useful diagnostic of this is provided by the right (left) weight of the operator, which is the total weight of $O(t)$ on Pauli strings that have their rightmost (leftmost) non-identity operator on site $i$, and act as the local identity on all sites to the right (left) of $i$:
\begin{align}
\rho_r(i,t)= \sum_{\substack{{\text{strings $\mc{S}$ with } }\\ \mathclap{\text{ rightmost non-}}\\ \mathclap{\text{identity on site $i$}}}} |a_\mc{S}(t)|^2.
\end{align}
The left weight $\rho_l(i,t)$ is defined analogously.  If $ O$ is initially local on site $j$ then $\rho_r(i,0) = \rho_l(i,0) = \delta_{ij}$. The conservation of operator norm implies that $\sum_i \rho_{r/l}(i,t)=1$, which gives $\rho$ the interpretation of an emergent local conserved `density" for the right/left fronts of the spreading operator. Refs.~\cite{opspreadAdam, opspreadCurt} showed that the (hydro)dynamics for $\rho_R(i,t)$ is governed by a biased diffusion equation, corresponding to fronts that propagate ballistically but broaden diffusively. Thus, as the operator spreads, $\rho_r$ moves right at $v_{B,r}$ and $\rho_l$ moves left at $v_{B,l}$. 

In order to compare the propagation of the left- and right fronts, we look at  $\rho_r(x_r+1,t)$ and $\rho_l(L-x_l,t)$. Thus, $x_{r}$ and $x_l$ are distances from the initial operator located at the left/right ends respectively. Note that $i$ runs from 1 to $L-1$ because it is a label while $x$ runs from 0 to $L$ because it is a distance. Fig.~\ref{fig:Rweightpeakshape}(top) shows $\rho_r(x_r+1,t)$ and $\rho_l(L-x_l,t)$ at successive times for different spatial separations $x_{l/r}$ from the starting locations in a system of size $L=13$, clearly showing ballistically traveling operator fronts. Note that the weights at equivalent distances from the ends of the chain peak at later times for the left-moving wave,  clearly showing $v_{B,l}<v_{B,r}$. More quantitatively, we can extract $v_{B,l}$ and $v_{B,r}$ from these curves by obtaining the times at which $\rho_{r/l}$ reach half their maximum peak height for a given distance $x$ (denoted by crosses/pluses on Figure~\ref{fig:Rweightpeakshape},top), and fitting these to linear functions (Figure~\ref{fig:Rweightpeakshape}, bottom). This procedure gives $v_{B,r}=1.011 \pm 0.072$ and $v_{B,l}=0.791 \pm 0.062$, showing a clear asymmetry in the butterfly speeds in the two directions. 

As mentioned earlier, because of the nature of the three-site term in the Hamiltonian, the right/left weights exhibit an ``odd-even" effect where site 3 may peaks before 2, etc (also visible in Fig.~\ref{fig:colorplot}). It is possible to account for these by averaging judiciously, or by  looking at only alternate sites, which is why we only show even distances in Figure~\ref{fig:Rweightpeakshape}.

\subsection{Asymmetric butterfly speeds from OTOCs}
\label{sec:HamOTOC}
We now turn to a complementary measure of operator spreading, namely the out-of-time-ordered commutator $C(i,t)$, defined for $\sigma^z$ operators as: 
\begin{align}
C(i,t) & = \half \langle|[\sigma^z_j(t),\sigma^z_i(0)]|^2\rangle_{\beta=0}\nonumber\\
&= 1 - \frac{1}{2^{L}}\Re\;\Tr\;[\sigma^z_j(t)\sigma^z_i(0)\sigma^z_j(t)\sigma^z_i(0)]
\label{eq:otoc}
\end{align}
where $j$ is the site index of the initial operator, and the expectation value in the top row is with respect to a thermal ensemble at infinite temperature. 
If $i$ is away from the initial location of the operator at $j$, then the operators on the different sites initially commute and the OTOC is zero. As the operator spreads, the OTOC grows to become of order one inside a ballistically growing light-cone defined by the left and right butterfly speeds, and is exponentially small outside it. The OTOC is related to the commutator norm that appears in Lieb Robinson bounds for local quantum systems~\cite{Lieb72}, and the Lieb Robinson velocity $v_{LR}$ serves as an upper limit on the maximum $v_B$~\cite{RobertsSwingle} in any direction. Fig.~\ref{fig:colorplot} shows the OTOC for an operator initially at the center of the chain, and the lightcone is approximately demarcated by where $C(i,t)= .4$, illustrated by the black bars. The figure again visually shows $v_{B,r} > v_{B,l}$.  The light-cone in the figure is not strictly monotonic at early times because of the even-odd effects mentioned earlier.

\begin{figure}
	\includegraphics[width=\columnwidth]{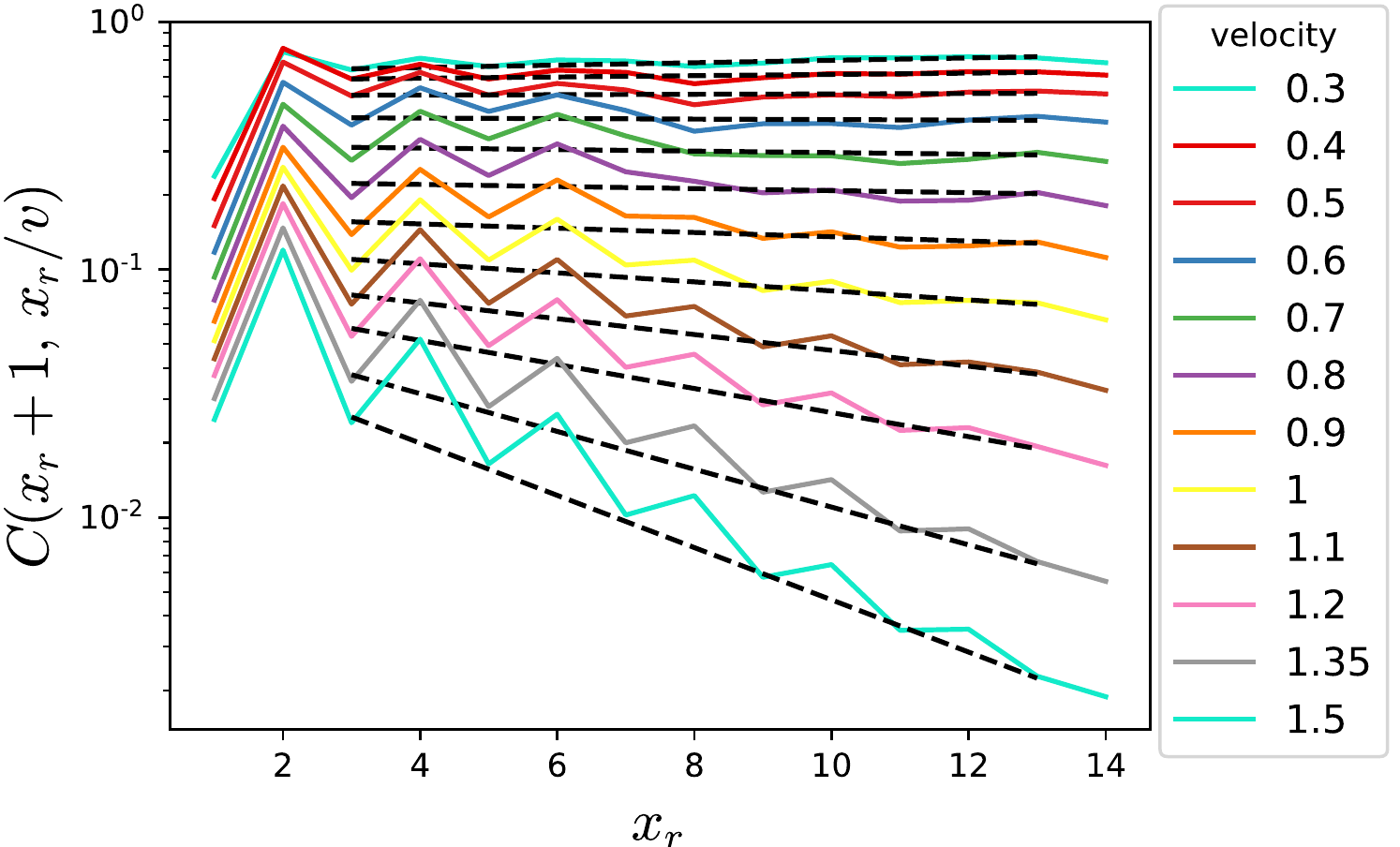}
	\includegraphics[width=\columnwidth]{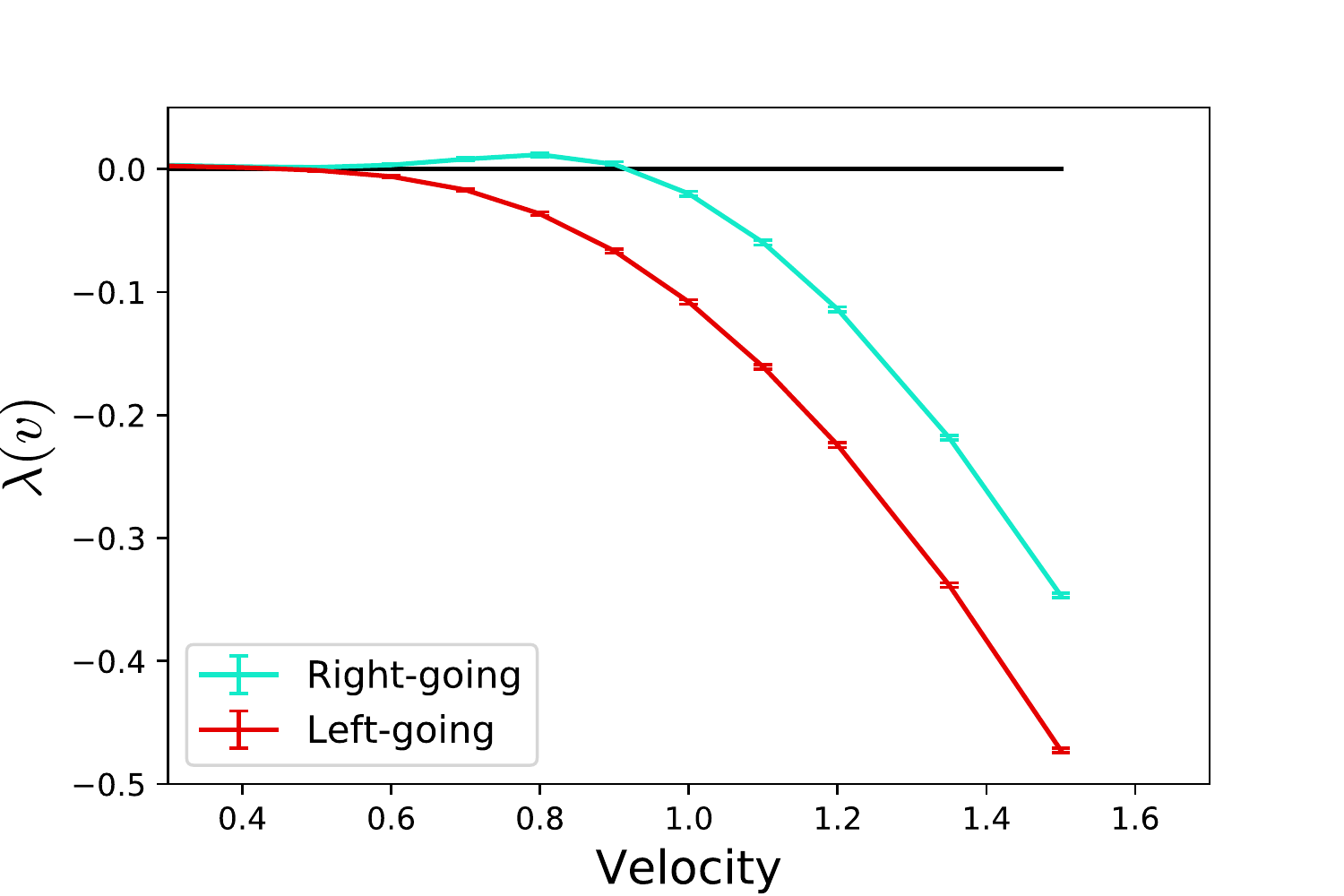}
	\caption{Velocity-dependent Lyapunov exponents extracted from the OTOC. The parameters are $L=15, h=0.35$. The top panel shows the forward (right propagating) OTOCs along rays at different velocities $v$ for a given disorder realization, with the initial operator at site 1. The data shows exponential decay, consistent with negative VDLEs for large $v$. VDLEs are obtained from the best fit line through alternate sites, with the slope equaling $\lambda(v)/v$. The lower figure shows the extracted $\lambda(v)$ averaged over 100 disorder realizations plotted against $v$, where the data for backward OTOCs is obtained via a similar method. Error bars on $\lambda(v)$ are obtained from the standard error of the mean. The figure clearly shows $\lambda_r(v)>\lambda_l(v)$ and, correspondingly, $v_{B,r}>v_{B,l}$.}
	\label{fig:vdle}
\end{figure}

The OTOC $C(i,t)$ is widely regarded as a diagnostic of chaos because, in many systems of interest, it shows an exponential growth with time from a value near zero to an order one number as the site $i$ enters the light-cone of the spreading operator, $C(i,t) \sim \epsilon e^{\lambda t}$ where $\lambda >0$ is a positive Lyapunov exponent~\cite{KitaevSYK, ShenkerStanfordButterfly,LocalizedShocks,CotlerRM,RobertsStanford,GuQiStanford,StanfordWeakCoupling,PatelDiffusiveMetal,ChowdhuryON,AleinerOTOC}. This is closely related to the exponential sensitivity of classically chaotic systems to small perturbations in initial conditions. However, an important point is that in quantum systems this exponential growth only takes place in systems which are in certain semiclassical or weakly coupled or large $N$ limits, and not in ``strongly-quantum" systems away from such limits~\cite{Khemani2018lambda, Swingle_otocMPS} --- such as strongly interacting thermalizing spin 1/2 chains which are the subject of this paper. Thus, no well-defined positive Lyapunov exponent exists in such strongly quantum systems. Nevertheless, Ref.~\cite{Khemani2018lambda} showed that one can still define \emph{velocity dependent} Lyapunov exponents (VDLEs) in these cases, and these can be used to provide a more detailed window into asymmetries in information propagation.  

The VDLEs $\lambda(v)$ are defined by looking at the growth or decay of OTOCs along \emph{rays} in spacetime at velocities $v$:
\begin{align}
	C(i, t) \sim e^{\lambda(v)t}\quad\text{for}\quad i = vt.
\end{align}
In all spatially local systems, $\lambda(v)$ is \emph{negative} for rays $v > v_B$ outside the lightcone defined by $v_{B, l/r}$, and smoothly approaches zero as $v\rightarrow{v_B^+}: \lambda(v) \sim -(v-v_B)^\alpha$~\cite{Khemani2018lambda}. Thus, even in strongly-quantum systems where it may not be possible to define positive Lyapunov exponents, one can define negative VDLEs which quantify the exponential \emph{decay} of information propagation outside the light cone. Note our convention where both butterfly speeds $v_{B, l/r}$ are defined to be positive and refer to the magnitudes of the velocities, even though the left side of the light-cone corresponds to negative velocities. 

As in the previous section, we measure information propagation to the right and left by setting the initial operator at $j=1$ and $j=L$ respectively. Once again, we measure the OTOC for positive distances $x = v t$ from the end sites for different positive $v$'s, understood to be the speed rather the velocity while considering backward propagation. Fig.~\ref{fig:vdle} shows the right-propagating OTOCs along rays at different speeds $v$ for a given disorder realization in a system of size $L=15$, showing the expected exponential decay for large $v$'s (left propagating curves are similar, not shown). There is a strong even-odd effect, so we only look at alternate sites to compute $\lambda(v)$. Note that the independent variable is distance in the top plot, so the slope of the best fit line on the semi-log plot is $\lambda(v)/v$. 

These $\lambda(v)'s$ so extracted are then averaged over disorder realizations and plotted against $v$ for both the right and left propagating OTOCs in the bottom panel of Fig.~\ref{fig:vdle}. Then, $v_{B, l/r}$ is the velocity at which $\lambda(v)$ smoothly goes to 0 for the left/right curves respectively. Note that the $\lambda(v)$ curves clearly show $\lambda(v)_r > \lambda(v)_l$ with $v_{B,r} > v_{B,l}$. These curves characterize the \emph{entire} region of spacetime outside the lightcone, and the difference between the left/right curves for all $v's$ illustrates that the asymmetry in information propagation is apparent everywhere in this region, not just at the edge of the lightcone. Thus, the VDLEs give more information about asymmetric information propagation than the left and right butterfly speeds alone. 

Estimating where the $\lambda(v)$ curves pass through zero, we see that $v_{B,l} \sim 0.6$ and $v_{B,r} \sim 0.9$. These are in the same ballpark as the velocities estimated from the right weight, but not exactly in agreement. It is reasonable to expect that for large enough sizes, there is a single speed for information propagation in every direction which agrees across these different methods, but it interesting to ask whether there can be different speeds diagnosed by different observables.

The data presented in Fig~\ref{fig:vdle} is for OTOCs involving $\sigma^z$ operators in a system of size $L=15$. The model \eqref{eq:Hammodel} conserves total $S^z$, so the computation of the OTOC reduces to a block diagonal form. For smaller blocks, we compute the OTOC directly using exact diagonalization, while for larger blocks we use canonical typicality to approximate the OTOC, as described in \cite{LuitzScrambling}. In this method, the OTOC is evaluated by using the second line in Eq.~\eqref{eq:otoc} and time-evolving randomly chosen pure stares drawn from the Haar measure. Thus, the data for larger blocks also involves an average over randomly chosen initial states. The error in using this method falls off exponentially with system size, and the number of pure states sampled is chosen to get a relative error less than $5\%$.

\section{Random staircase circuit models with asymmetric butterfly speeds} \label{sec:circ}

We now switch gears and discuss a different system that also displays asymmetric spreading, and can be made completely chiral in a certain limit. Instead of time-independent Hamiltonians, we will consider a class of random circuit models, called ``$n$-staircase" models, with a tunable parameter $n$ controlling the degree of asymmetry. While unitary circuit models with asymmetric information propagation are well known, one of our main messages in this section will be to demonstrate that the functional form of the ``entanglement generation function" which controls the coarse-grained entanglement and operator dynamics, defined in Ref.~\cite{Jonay} and below, can be varied systematically in the $n$-staircase models while still respecting certain general constraints this function must obey \cite{Jonay}. This function also controls operator spreading, and gives yet another way to probe asymmetric information transport. 

The setting is again a spin chain of length $L$, but now the local Hilbert space dimension on each site is $q \geq 2$. The model can be mapped to an efficiently simulable classical model in the $q\rightarrow{\infty}$ limit. At each discrete time step, unitary gates act on pairs of consecutive sites. Each pair of sites is specified by a bond index and, in contrast to the previous section, we will refer to sites with index $i$ and bonds with index $x$. Each two-site gate is chosen randomly and independently from the uniform Haar measure. In addition, unlike the circuits in Figure~\ref{fig:circuits}, the architecture of this circuit is also random so that a random gate acts on a randomly chosen bond at every time step, but with certain correlations which encode the ``staircase" architecture indexed by the staircase size $n$. Each ``$n$-stair" is a contiguous string of $n$ random gates acting on bonds $x$ through $x+n-1$ in succession; the staircase circuit consists of $n$-stairs placed at random bonds $x$ at each time step.  For $n=1$ this is just the uncorrelated random circuit whose entanglement dynamics were studied in \cite{AdamCircuit1}, but large $n$ results in more asymmetric circuits. Fig.~\ref{fig:circuits}(c) is a regular Floquet circuit build from 3-stairs without any randomness in the circuit architecture. 

We note that staircases were introduced in~\cite{opspreadAdam}, but \cite{opspreadAdam} included both left- and right-facing staircases with the goal of engineering an arbitrarily high ratio of $v_B$ in relation to $v_E$, the entanglement velocity. In addition, Ref.~\cite{Jonay} pointed out that preferentially including more of one type of staircase will lead to asymmetric butterfly speeds. In our circuits, with only right-staircases, we achieve an arbitrarily high ratio of $v_{B,r}$ in relation to $v_E$.

% \vedika{We will be interested in circuits with infinite $L$. When $n$ is small we can extract the behavior from systems with finite $L>>n$. For large $n$ we can first take $L\to\infty$ and then $n\to\infty$, or set $n=L$ and take them to $\infty$ together. The behavior does not depend on the order of limits, but it is easier to reason about the circuits using the second procedure.

The structure of this section is as follows. Subsection~\ref{sub:entropy} describes the entanglement growth/generation function and how this function encodes the butterfly velocity for operator spreading. In Subsection~\ref{sub:classical} we show that the limit $q\to\infty$ results in efficiently simulable classical dynamics for the entropy. Then, in Subsection~\ref{sub:asym} we explore the left and right butterfly speeds $v_{B, l/r}$ for these circuits. The transport is symmetric for $n=1$, and completely unidirectional for $n=\infty$. Although the model is not solvable for intermediate $n$, we provide an approximation to the entropy growth function that is correct at $n=1, \infty$. 

\subsection{Entropy in random circuits} \label{sub:entropy}
For our one dimensional system of interest, let $S(x,t)$ be the bipartite von Neumann entanglement entropy across bond $x$ at time $t$. Building on work by Nahum et. al.~\cite{AdamCircuit1}, Ref.~\cite{Jonay} recently presented an effective coarse-grained ``hydrodynamical" description for the entanglement dynamics which assumes that, to leading order, the local increase in entanglement is set by an entanglement production rate $\Gamma(s)$, which  depends on the local entanglement gradient $s = \frac{\partial S}{\partial x}$: 
\begin{align}
\frac{\partial S(x,t)}{\partial t} = s_{\rm eq} \Gamma\left(\frac{\partial S(x,t)}{\partial x}\right) + \cdots
\end{align}
where $s_{\rm eq}$ is the equilibrium entropy density in thermal equilibrium. This function $\Gamma(s)$, called the entanglement growth or generation function, governs the entanglement and operator dynamics and will be a central object of study for us in this section, providing yet another measure to diagnose asymmetric information spreading.   

For a system in thermal equilibrium, the entanglement production rate must go to zero. In equilibrium, $S(x,t)$ has a ``pyramid" shape $S(x,t) = s_{\rm eq} {\rm min} \{x, L-x\}$ in a system of length $L$, so that $\Gamma(s_{\rm eq}) = \Gamma(-s_{\rm eq})=0$, while for $|s| < s_{\rm eq}$, $\Gamma(s)$ is positive. In can be shown that this function also determines the dynamics of operator spreading, and the butterfly speeds are given by the derivative $\Gamma'(s)|_{s_\text{ext}}$, where $s_\text{ext}$ is one of the extremal entropy densities, $s_{eq}$ or $-s_{eq}$~\cite{Jonay}: 
\begin{align}
v_{B,l}=s_{\rm eq} \left.\frac{\partial \Gamma(s)}{\partial s}\right|_{s=-s_{eq}}, \quad 
    v_{B,r}=-s_{\rm eq} \left.\frac{\partial \Gamma(s)}{\partial s}\right|_{s=s_{eq}} 
    \label{eqn:vbGamma}
\end{align}
The signs ensure that both velocities are positive.
It follows that any $\Gamma(s)$ with asymmetry at the endpoints will have asymmetric butterfly velocities. Finally, for completeness, we note that the entanglement speed is given by $v_E = \Gamma(0)$, and $\Gamma(s)$ must be a convex function which implies that $v_B \geq v_E$. 

\subsection{Classical dynamics in the $q\to\infty$ limit} 
\label{sub:classical}
While solving the evolution of $S(x,t)$ in full generality is nearly impossible, there are certain limits in which this analysis is simplified~\cite{AdamCircuit1}. If a gate acts on bond $x$, it can increase the bipartite entanglement entropy $S(x)$, up to a constraint $|S(x + 1) - S(x)| \le 1$ which follows from subadditivity~\footnote{Logarithms for defining the entropy are taken in base $q$}. In the $q\to\infty$ limit, a Haar-random gate will, with probability 1, maximally increase the entanglement across the bond it acts on~\cite{AdamCircuit1}. Given the previous constraint, this means that if a gate acts at bond $x$ at time $t$, then~\cite{AdamCircuit1} $$S(x, t+1) = \min\left\lbrace S(x-1,t), S(x+1,t)\right\rbrace+1.$$ For the remainder of this paper we will use the $q\to\infty$ limit, while $n$ will still be variable.

At this point, all quantum effects leave the system, and the information dynamics are purely classical. This means it is possible to simulate the circuit without diagonalizing any Hamiltonians or unitary operators. It suffices to consider integer-valued $S(x)$ with $\Delta S(x) \equiv S(x)-S(x-1)=\pm 1$ for all $x$. Any other state, with either non-integer $S(x)$ or flat steps, will approach one with these characteristics~\cite{AdamCircuit1}. A state of this form can be described as a series of ``up" and ``down" steps at each site corresponding to $\Delta S = +1$ and $-1$ respectively, denoted $u$ and $d$. If a gate falls on bond $x$, it adds two units of entropy to $S(x)$ iff the step before is $d$ and the step after is $u$ -- that is, entanglement is locally generated only at places where $S(x)$ has a local minima. This classical evolution is deterministic and can be easily simulated.
Since individual circuits have deterministic behavior, we average over circuit architecture (the random placement of $n$-stairs).

\begin{figure}
	\includegraphics[width=\columnwidth]{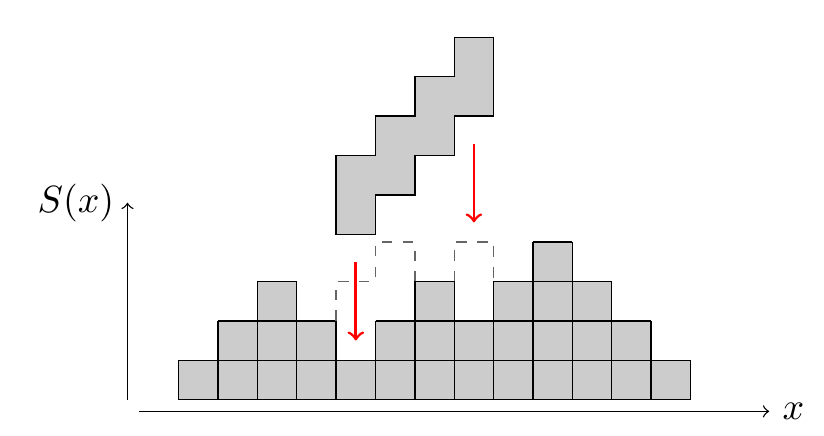}
	\caption{A 4-staircase falling on an example entropy function. Note that each gate raises $S(x)$ by 2 iff $S(x)$ is a local minimum when that gate falls. So the second gates does in fact hit a local minimum because it acts after the first gate.}
	\label{fig:stairs}
\end{figure}

Figure \ref{fig:stairs} illustrates the evolution of the entropy function for a single application of a 4-stair. The stair consists of 4 individual gates. Each gate has height 2 because if it produces entropy, it produces 2 units. The shaded profile is the initial $S(x)$, while the dashed line shows $S(x)$ after the $n$-stair falls. The first, second, and fourth gates were productive while the third was not.
It is already possible to see the origin of the asymmetry. The 4-staircase can only be perfectly effective (produce 8 units of entropy) if it hits the microstate ${d,u,u,u}$. Any other state will result in the production of less entropy, if any. Since this microstate has positive slope, it is more likely to be found when the entropy gradient is larger.

To set a time scale for the system, we need a rate at which gates are applied. The gate rate $\gamma$ is defined as the number of \emph{gates} per unit time, not staircases. This means that as if every gate is productive, the entropy growth rate is $\Gamma=2\gamma$. The rate at which complete staircases fall is $\gamma/n$.

\subsection{Asymmetric $v_B$} \label{sub:asym}

Despite the deterministic evolution of these circuits, they cannot be solved for finite $n>1$. This is due to correlations that arise in the up and down steps of $S(x)$. If these steps were uncorrelated, then a general state could be described solely by the (average) local entropy gradient $s$. At any bond the probability of $u$ would be $(1+s)/2$ and the probability of $d$ would be $(1-s)/2$,  On the other hand, correlations make the probability of an $u$ dependent on the surrounding steps. 

There are no correlations in the $n=1$ model, so we can exactly solve the entropy growth function.
Since a gate is productive only at a local minimum, \emph{i.e} the microstate $\{d,u\}$, the probability that a randomly placed gate is productive is $(1+s)(1-s)/4$. Then, the entropy growth rate is the gate rate, times the probability of entropy production, times the entropy produced per gate:
\begin{align}
    \Gamma_1(s)=\gamma\frac{(1+s)(1-s)}{4}2 = \gamma\frac{1-s^2}{2}
\end{align}
For larger $n$ we can perform a similar analysis. Although we know correlations will affect the growth rate, we hope the effect is small. 

Consider next the case of $n=2$ with 2-stairs consisting of one gate acting at bond $x$ and one at bond $x+1$. The entropy production of these gates is affected by the slope between the two bonds and the slopes on either side. There are 8 possible configurations of those three slopes, but only 4 result in entropy growth, as shown in table~\ref{tab:2stair}. Weighting each configuration by its probability and the entropy generated, and then multiplying by the staircase rate $\frac{\gamma}{2}$, the growth rate, with correlations neglected, is:
\begin{align}
\Gamma_2(s) 
% &= \frac{\gamma}{2} 4\frac{1+s^2}{4}\frac{1+s}{2} + \frac{\gamma}{2}
% 	2\frac{1-ms2}{4}
% 	\left(\frac{1-s}{2} + \frac{1-s}{2} + \frac{1+s}{2}\right) \nonumber\\
&= \frac{\gamma}{2}\frac{1-s^2}{2}\frac{5+s}{2}, \label{eqn:2rate}
\end{align}
We can interpret the factor $\frac{1-s^2}{2}\frac{5+s}{2}$ as the average entropy produced by each staircase. The second factor provides the asymmetry. Of course, the repeated application of the 2-stairs will lead to a buildup of correlations between $u,d$ patterns in time, and this is the piece of physics that is ignored by our analysis.  

\begin{table}
	\centering
	\begin{tabular}{ccc}
		Initial $\to$ Final 
		Configuration        & Probability         & Productivity\\
		$d\,u\,d\to u\,d\,d$ & $\frac{1-s}{2}\frac{1+s}{2}\frac{1-s}{2}$ & 2\\
		$d\,u\,u\to u\,u\,d$ & $\frac{1-s}{2}\frac{1+s}{2}\frac{1+s}{2}$ & 4\\
		$d\,d\,u\to d\,u\,d$ & $\frac{1-s}{2}\frac{1-s}{2}\frac{1+s}{2}$ & 2\\
		$u\,d\,u\to u\,u\,d$ & $\frac{1+s}{2}\frac{1-s}{2}\frac{1+s}{2}$ & 2
	\end{tabular}
	\caption{The four configurations that result in entropy growth for 2-stairs, the relative proportions of the initial states assuming an uncorrelated entropy distribution, and the growth in entropy generated by a 2-stair falling on that configuration. The four configurations that do not result in entropy growth are $u\,u\,u, d\,d\,d, u\,d\,d,$ and $u\,u\,d$.}
	\label{tab:2stair}
\end{table}

We can determine the growth rate for arbitrary length stairs through a recursive relationship. Consider a staircase made of $n$ gates. Like in the $n=2$ case, its growth rate will be proportional to the staircase rate $\frac{\gamma}{n}$ multiplied by the average entanglement generated by each staircase, so we can write
\begin{align}
\Gamma_n(s) = \frac{\gamma}{n}R_n(s), \label{eqn:growthrateR}
\end{align}
where $R_n(s)$ is the average entropy production of an $n$-stair. To find an equation for $R_n(s)$, note that the first $n-1$ gates have the same entanglement production as the $(n-1)$-stair. All final states of the $(n-1)$-stair end in a down slope, so the $n$th gate will produce another 2 units of entropy if the last step is $u$. However, if all $n+1$ initial steps are $u$ no entanglement is generated. 

This behavior is captured by the recursive formula
\begin{align}
R_n(s) = R_{n-1}(s)+2\frac{1+s}{2} - 2\left(\frac{1+s}{2}\right)^{n+1}, \label{eqn:raterecur}
\end{align}
along with the base case $R_0(s)=0$. The solution is
\begin{align}
\Gamma_n(s) = \frac{\gamma}{n}\frac{1+s}{1-s}\bigg(
	(1+s)&\left[\left(\frac{1+s}{2}\right)^n-1\right]\nonumber \\
	&+n(1-s)\bigg). \label{eqn:growthrate}
\end{align}
Then, from Eqn.~\ref{eqn:vbGamma}, $v_{B,l}=\gamma$ while $v_{B,r}=\half\gamma(n+1)$, where we've used $s_{eq}=1$ because logarithms are taken base $q$.
This produces successively more asymmetric butterfly velocities as $n$ increases. Note also that this analysis also assumes that the ``local" coarse grained entropy gradient is well-defined on length scales longer than $n$, which is not a problem for small $n$.  

\begin{figure}
    \includegraphics[width=\columnwidth]{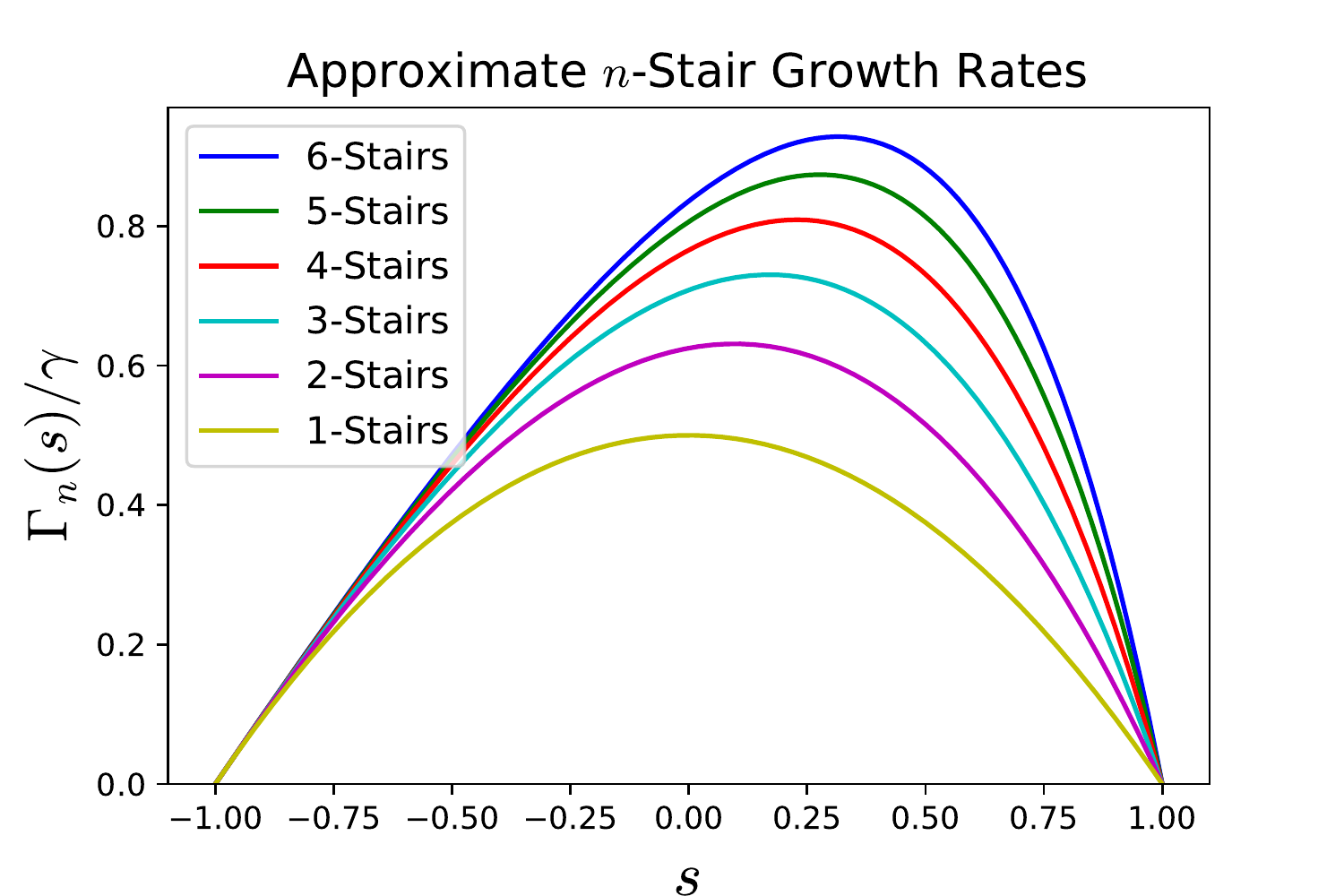}
	\includegraphics[width=\columnwidth]{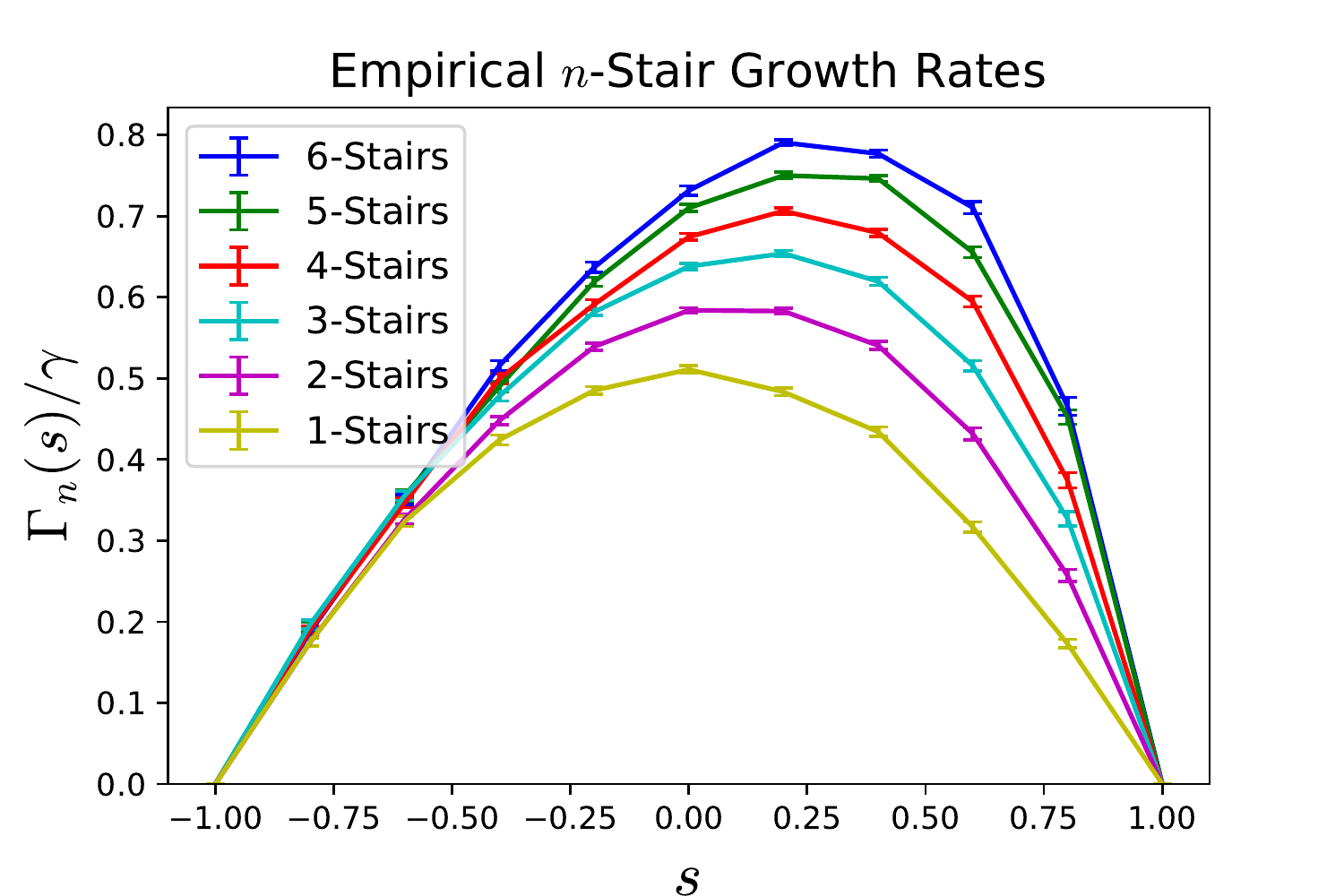}
	\caption{Approximate and empirical entanglement growth rate functions $\Gamma(s)$ as a function of the entropy gradient $s$ for $n$-stair circuits. The approximate growth rate~\eqref{eqn:growthrate} ignores correlations and consistently overestimates, but captures the qualitative trend. The right/forward and left/backward butterfly velocities are the slopes of these curves at their endpoint, indicating that while $v_{B,l}$ stays constant, $v_{B,r}$ increases. All growth rates were calculated using the classical simulation described in the text for a 100-site spin chain with periodic boundary conditions allowing for an average slope $s$. Rates were calculated from the application of 2,000 gates total, or 20 per site, averaged over the last 80\% of the gates in order to build up correlations, then averaged over 100 trials. }
	\label{fig:compareRates}
\end{figure}

The question then becomes, how much of an effect do the correlations have? 
For small $n$, we can simulate the classical dynamics numerically for finite $L>>n$. This will include all correlations ignored by \eqref{eqn:growthrate}. For the growth rate curves of $n$-stair circuits for $n\le 6$ see Fig.~\ref{fig:compareRates} (bottom). 
The asymmetry is evident for all $n>1$, and the asymmetry continues to increase as $n$ increases. The growth rates look qualitatively similar to the approximate, analytically derived rates ignoring correlations~\eqref{eqn:growthrate}, shown in Figure~\ref{fig:compareRates} (top), although they are smaller overall. 
	
As $n$ increases, exact simulation becomes difficult. 
However, luckily, as $n$ becomes very large (for infinite $L$) or approaches the size of the system (for finite $L$), the correlations again become unimportant. To see this, we show that the predicted growth rate is correct at various $s$ in this limit. The predicted growth rate in this limit is $\Gamma_\infty(s)=\gamma(s+1)$~\eqref{eqn:growthrate}. The butterfly velocities corresponding to this are $v_{B,l}=\gamma$ and $v_{B,r}=\infty$, corresponding to perfect chiral transport. 

Consider $s = -1, 0,$ and 1 for $n=L$ stairs. In this case, $s$ refers to the average entanglement gradient across the entire system: $sL = S(x+L)-S(x)$. At exactly $s=\pm1$ there will be no growth, so we consider an entropy function with a single up or down step before sending $L\to\infty$.

Near $s=-1$, the entropy profile consists of all down steps with a single $u$. Then the circuit generates entanglement once every time a staircase falls. We determine $\Gamma_\infty(-1)=\gamma/L$, which approaches 0 as $L$ becomes large.

In the $s = 0$ case, after a gate falls between sites $i$ and $i + 1$, $s_{i+1}$ will be a down slope regardless of whether the gate generated entanglement. Then the next gate falls across sites $i + 1$ and $i+2$. At site $i+2$ $s_{i+2} = u$ with probability $\half$, so on average $\half$ of the gates produce 2 units of entanglement and $\Gamma_\infty(0)=\gamma$. 

At near-maximal slope $s=1$, nearly all slopes are $u$, except at the site to the right of the most recent gate. Then the next gate falls at a local minimum with probability 1, and all gates produce 2 units of entanglement, so $\Gamma_\infty(1)=2\gamma$.

Because these rates match the predicated rate, we know it is exact at $s=0, \pm1$. From convexity, the only possible function is then $\Gamma_\infty(s)=\gamma(s+1)$. This shows that our approximation again becomes exact at $n=\infty$ and the system achieves chiral transport. The butterfly velocities are $v_{B,l}=\gamma$ and $v_{B,r}=\infty$. Although we do not know the exact behavior for $1<n<\infty$, we know it interpolates between symmetric and completely asymmetric behavior.

\section{Discussion}
In summary, we have constructed and studied  two  models  with  asymmetric butterfly speeds using a variety of complementary measures for the propagation of quantum information. The first model was a local, time-independent Hamiltonian, while the second was a class of ``staircase" circuits with a tunable parameter, capable of interpolating from symmetric spreading to completely chiral information propagation. The degree of asymmetry in the models considered is limited by notions of locality. In time-independent Hamiltonian systems, locality is measured by the range of interactions, while in Floquet circuits it is related to the depth or period of the circuit. In unitary circuits that are random in space and time, a notion of locality is encoded by the extent of spatial correlations in the circuit architecture, such as in the staircase models presented here.  

The Hamiltonian model presented here was quite general, with several tuning parameters. For example, it might be interesting to study how the asymmetry in butterfly speeds evolves as the system approaches the many-body localization transition where the butterfly speeds are zero and information propagation is logarithmic in time. Separately, another interesting direction of research could entail finding the most asymmetric Hamiltonians possible for a given range of interactions. 

It would also be interesting to study how asymmetries in various measures of particle and information transport are intertwined with each other. For example, must it be the case that different quantities like the transport of conserved densities (like energy of particles), the spreading of quantum entanglement, and the growth of local operators all inherit signatures of asymmetry, or is it possible to disentangle these?

\emph{Note Added---} The majority of these results were presented in the senior thesis of the lead author~\cite{CharlieThesis} in May 2018. While we were completing this manuscript, a study appeared also showing asymmetric information transport in local, time-independent Hamiltonians~\cite{GorshkovAsymm}, but the asymmetry in that work is derived from particle exchange statistics. By contrast, our paper derives derives models with asymmetric information transport while keeping particle statistics fixed, and is thus complementary to~\cite{GorshkovAsymm}.

\emph{Acknowledgements---}
We thank Adam Nahum, Cheryne Jonay and David Luitz for helpful discussions. VK is supported by the Harvard Society of Fellows and the William F. Milton Fund.

\bibliography{global}

\end{document}